\documentclass[onecolumn,secnumarabic,amssymb, nobibnotes, aps, prd]{revtex4}
\usepackage{amsmath} \usepackage{amssymb}
\usepackage{graphicx} 
\usepackage{epstopdf}
\usepackage{amsmath}
\usepackage{amsmath,amssymb,amsthm,amsfonts,mathrsfs,bm,verbatim}
\usepackage{graphicx,subfigure}

\newcommand{\bea}{\begin{eqnarray}}
\newcommand{\eea}{\end{eqnarray}}

\begin{document}

\title{
Analysis on superradiant stability of BTZ Black Hole }%

\author{Wen-Xiang Chen$^{a}$}
\affiliation{Department of Astronomy, School of Physics and Materials Science, GuangZhou University, Guangzhou 510006, China}
\author{Yao-Guang Zheng}
\email{hesoyam12456@163.com}
\affiliation{Department of Astronomy, School of Physics and Materials Science, GuangZhou University, Guangzhou 510006, China}

\begin{abstract}
This paper adds a new variable y($\mu^{\prime}=y(\omega+m N^{\phi})$) to extend the results of the classic paper. We exploit the properties of curve integrals. When y is greater than a certain limit, the effective potential of the equation has no pole, then there is no potential well outside the event horizon, when $\sqrt{2(m^2)}/{ r^2_+}< \omega < m\varOmega_H$,so  the BTZ black hole was superradiantly stable at that time.

 \text { KEYwords: } Superradiance; BTZ black hole; Curvilinear integral

\end{abstract}

\maketitle

\section{Introduction}
The BTZ black hole is a special $(2+1)$-dimensional black hole, and the metric of the BTZ black hole can be obtained by solving the $(2+1)$-dimensional Einstein-Maxwell equation with a negative cosmological constant. BTZ black holes have special space-time structure and relatively simple metric, so it is very suitable to be used to study the relevant properties of black holes.

Whether it is the classical characteristics of black holes or quantum characteristics, people are fascinated. For a long time, people have been eager to find a lower-dimensional simulated black hole, so as to study the key properties of black holes, and in this way, the unnecessary complicated processes that exist in the study of high-dimensional black holes can be omitted. Fortunately, BTZ black holes are good candidates. The metric of a BTZ black hole can be obtained from the standard Einstein-Maxwell equation with a negative $(2+1)$-dimensional cosmological constant\cite{1}.

The phenomenon of superradiance finds a universal realization in black hole physics: the ability to extract energy and angular momentum from a rotating BH is a rigorous mathematical treatment. This extraction is mediated by a low-frequency boson mode that is amplified and exponentially growing as it scatters by a spinning black hole, and becomes unstable if trapped near the black hole. Over the past few decades, a great deal of work has been devoted to understanding this instability when the trapping is enforced by the mass term of the boson field\cite{2,3,4,5,6,7,8}.

The black hole no hair theorem was first proposed by Wheeler in 1971 and proved in 1973 by Stephen Hawking\cite{1}, Brandon Carter and others. The development of black hole thermodynamics in the 1970s applied the fundamental laws of thermodynamics to the theory's place in the field of general relativity, and strongly hinted at the deep and fundamental relationship between general relativity, thermodynamics, and quantum theory.
The stability of black holes is an important topic in black hole physics. Regge and Wheeler\cite{2} proved that spherically symmetric Schwarzschild black holes are stable under perturbation. The huge effects of superradiation complicate the stability of spinning black holes. The superradiant effect occurs in both classical and quantum scattering processes. When a boson wave hits a rotating black hole, if certain conditions are met, the rotating black hole is likely to be as stable as a Schwarzschild black hole\cite{1,2,3,4,5,6,7,8,9}
\bea\label{src}
\omega< m\varOmega_H+q\varPhi_H, \varOmega_H=\frac{a}{r_+^2+a^2}
\eea
Where $q$ and $m$ are the charge number and azimuthal quantum number of the incident wave, $\omega$ represents the wave frequency, $\varOmega_H$ is the angular velocity of the black hole event horizon, $\varPhi_H$ is the electromagnetic potential of the black hole event horizon. If the frequency range of the wave is superradiant, the wave reflected by the event horizon is amplified, which means that when the incoming wave is scattered, the wave extracts rotational energy from the spinning black hole. According to the black hole bomb mechanism \cite{1,2,3,4,5,6,7,8,9} proposed by Press and Teukolsky, if a mirror black hole is placed between the event horizon and the outer space of the black hole, the amplified wave would bounce back and forth between the mirror and the black hole and grow exponentially, causing the black hole's superradiant instability.

This paper adds a new variable y($\mu^{\prime}=y(\omega+m N^{\phi})$) to extend the results of the classic paper. We exploit the properties of curve integrals. When y is greater than a certain limit, the effective potential of the equation has no pole, then there is no potential well outside the event horizon, when $\sqrt{2(m^2)}/{ r^2_+}< \omega < m\varOmega_H$,so  the BTZ black hole was superradiantly stable at that time.

\section{Description of the system and superradiance}

We know this method of horizon thermodynamics in (2+1)-dimensional Einstein gravity to explain how it works. Consider the space-time of a BTZ black hole, the geometry of which is given by\cite{5,6,7}
\begin{equation}
d s^{2}=-f(r) d t^{2}+f^{-1}(r) d r^{2}+r^{2}\left(d \phi+N^{\phi} d t\right)^{2}
\end{equation}
\begin{equation}
\begin{aligned}
&f(r)=\frac{r^{2}}{l^{2}}-8 G_{N} M+\frac{J^{2}}{4 r^{2}} \\
&N^{\phi}=-\frac{J}{2 r^{2}}
\end{aligned}
\end{equation}
\begin{equation}
r_{\pm}^{2}=4 G_{N} M l^{2}\left[1 \pm \sqrt{1-\left(\frac{J}{8 G_{N} M l}\right)^{2}}\right]
\end{equation}

Because Newton's gravitational constant $\mathrm(G)$ is related to the dimension of space-time, the third chapter considers four-dimensional space-time, using the geometric unit system, take $\mathrm(G)=1$. But for three-dimensional space-time, it’s convenient , usually take $8 \mathrm{G}=1$. In this chapter, all $\mathrm{G}$ are written out, and to distinguish the following Gibbs function, this chapter records the Newton gravitational constant G as $G_{N}$ .

We mainly consider scalar fields. The scalar field $\varphi$ satisfies the Klein-Gordon equation\cite{9}
\begin{equation}
\left(\square-\lambda R-\mu^{2}\right) \varphi=0,
\end{equation}
where $\lambda$ is the conformal coupling constant, which disappears when the scalar field is minimally coupled to the background metric. The curvature is given by $R=-\frac{6}{\ell^{2}}$, and $\mu$ can be regarded as the field mass. $\square$ represents the D'Almbert operator, given by\cite{9}
\begin{equation}
\square \equiv \frac{1}{\sqrt{-g}} \partial_{\mu}\left(\sqrt{-g} g^{\mu \nu} \partial_{\nu}\right)
\end{equation}
Then the equation  can be rewritten as
\begin{equation}
\frac{1}{\sqrt{-g}} \partial_{\mu}\left(\sqrt{-g} g^{\mu \nu} \partial_{\nu} \varphi\right)-{\mu^{\prime}}^{2}\varphi=0
\end{equation}
where ${\mu^{\prime}}^{2}=\mu^{2}+\lambda R$. Make the ansatz
\begin{equation}
\varphi(t, r, \phi)=e^{i(\omega t+m \phi)} \frac{R(r)}{\sqrt{r}},
\end{equation}
where $\omega$ is the frequency or energy and $m$ is the axial quantum number or angular momentum.

The radial wavefunction $R(r)$ satisfies the following equation
\begin{equation}
\frac{d^{2} R(r)}{d r^{* 2}}+V_{\mathrm{eff}} R(r)=0,
\end{equation}
where $r^{*}$ is the tortoise coordinate, and the effective potential $V_{\text {eff }}$ can be written as
\begin{equation}
V_{\mathrm{eff}}=\left\{\left(\omega+m N^{\phi}\right)^{2}-f(r)\left[\frac{m^{2}}{r^{2}}+\mu^{\prime 2}+\frac{1}{2} \frac{d}{d r}\left[\frac{f(r)}{r^{\frac{3}{2}}}\right]+\frac{f(r)}{2 r^{\frac{2}{5}}}\right]\right\}
\end{equation}
where $r^{*}$ denotes the tortoise coordinate
\begin{equation}
r^{*}(r)=\int f^{-1}(r) d r=\frac{1}{2\left(r_{+}^{2}-r_{-}^{2}\right)}\left[r_{+} \log \frac{\left|r-r_{+}\right|}{r+r_{+}}-r_{-} \log \frac{\left|r-r_{-}\right|}{r+r_{-}}\right] .
\end{equation}

We know that the frequency
\begin{equation}
\omega_{2}=\sqrt{\left(\omega+m N^{\phi}(\infty)\right)^{2}-\mu^{\prime 2}}=\sqrt{\omega^{2}-\mu^{\prime 2}} .
\end{equation}
For the radial wavefunction $R(r)$, we have
\begin{equation}
\begin{cases}R_{1}\left(r^{*}\right)=e^{-i \omega_{2} r^{*}}+\mathcal{R} e^{i \omega_{2} r^{*}}, & r \rightarrow \infty \\ R_{2}\left(r^{*}\right)=\mathcal{T} e^{-i \omega_{1} r^{*}}, & r \rightarrow r_{+}\end{cases}
\end{equation}
where $\mathcal{R}$ and $\mathcal{T}$ are the reflection and the transmission coefficients, respectively. One can easily obtain the following expression from
\begin{equation}
W=R_{1} \frac{d R_{2}}{d r^{*}}-R_{2} \frac{d R_{1}}{d r^{*}},
\end{equation}
where $W$ is the Wronskian. On the other hand, the complex conjugate $R_{1,2}^{*}$ is another linearly independent solution. Evaluating the above equation, we find
\begin{equation}
|\mathcal{R}|^{2}=1-\frac{\omega_{1}}{\omega_{2}}|\mathcal{T}|^{2}
\end{equation}
Therefore,
\begin{equation}
0<\omega<m \Omega_{H}
\end{equation}
then the reflection coefficient $|\mathcal{R}|^{2}>1$.

\begin{equation}
 V'( r\rightarrow +\infty )
 \rightarrow \frac{(2-y^2)}{r^2}+{\cal O}( \frac{1}{r^3}) ,
\end{equation}
The derivative of the effective potential has to be negative in order to satisfy the no trapping well condition,
\begin{equation}
2-y^2<0.
\end{equation}

\section{The limit $y$ of the incident particle under the superradiance of BTZ black holes}
We will now clearly show that the Schrodinger-like equation determines the radial function behavior of the spatially bounded non-minimum coupled-mass scalar field configuration of the BTZ black hole space-time, suitable for WKB analysis of large masses. In particular, the Schrodinger-like standard two of the radial equation Order WKB analysis yields the well-known discrete quantization condition(Here we have used the integral relation $\int_{0}^{1} d x \sqrt{1 / x-1}=\pi / 2$.),when $V( r\rightarrow +\infty ),{\mu^{\prime}}=1/(n+{1\over2})$, \cite{10,11,12}
\begin{equation}
\int_{(y^2)_{t-}}^{(y^2)_{t+}}d(y^2)\sqrt{\omega ^2-V(y;M,m,{\mu^{\prime}},\alpha1)}=\big(n+{1\over2}\big)\cdot\pi{\mu^{\prime}} /{{2}}=\pi/2
\ \ \ ; \ \ \ \ n=0,1,2,...\  .
\end{equation}
The two integration boundaries $\{y_{t-},y_{t+}\}$ of the WKB
formula are the classical turning points [with
$V(y_{t-})=V(y_{t+})=0$] of the composed
charged-black-hole-massive-field binding potential .
The resonant parameter $n$ (with $n\in\{0,1,2,...\}$) characterizes
the infinitely large discrete resonant spectrum
$\{\alpha_n({\mu^{\prime}},m)\}_{n=0}^{n=\infty}$ of the black-hole-field
system.

Using the relation  between the radial coordinates $y$ and $r$, one can
express the WKB resonance equation  in the form
\begin{equation}
\int_{r_{t-}}^{r_{t+}}dr{{\sqrt{-V(r;M,m,{\mu^{\prime}},\alpha)}}\over{f(r)}}=\big(n+{1\over2}\big)\cdot\pi\
\ \ \ ; \ \ \ \ n=0,1,2,...\  ,
\end{equation}
determine the radial turning points $\{r_{t-},r_{t+}\}$ of the composed black-hole-field binding potential.

We set
\begin{equation}
x\equiv {{r-r_{\text{+}}}\over{r_{\text{+}}}}\ \ \ \ ; \ \ \ \ \tau\equiv {{r_+-r_-}\over{r_+}}\  ,
\end{equation}
in terms of which the composed black-hole-massive-field interaction term
has the form of a binding potential well,
\begin{equation}
V[x(r)]=-\tau\Big({{\alpha(m^2)}\over{r^4_+}}-{\mu^{\prime}}^2\Big)\cdot x +
\Big[{{\alpha(m^2)(5r_+-6r_-)}\over{r^5_+}}-{\mu^{\prime}}^2\big(1-{{2r_-}\over{r_+}}\big)\Big]\cdot x^2+O(x^3)\  ,
\end{equation}
in the near-horizon region
\begin{equation}
x\ll\tau\  .
\end{equation}

From the near-horizon expression  of the
black-hole-field binding potential, one obtains the dimensionless
expressions
\begin{equation}
x_{t-}=0\
\end{equation}
and
\begin{equation}
x_{t+}=\tau\cdot{{{{\alpha (m^2)}\over{r^4_+}}-{\mu^{\prime}}^2}\over{{{\alpha (m^2)(5r_+-6r_-)}\over{r^5_+}}-{\mu^{\prime}}^2\big(1-{{2r_-}\over{r_+}}\big)}}\
\end{equation}
for the classical turning points of the WKB integral relation .

We find that
our analysis is valid in the regime  below($\alpha1$ corresponds to the transformation of y )
\begin{equation}
\alpha\simeq{{{\mu^{\prime}}^2r^4_+}\over{(m^2)}}  ,\alpha1\simeq\sqrt{{{{\mu^{\prime}}^2r^4_+}\over{(m^2)}}}\
\end{equation}
in which case the near-horizon binding potential and its outer
turning point can be approximated by the remarkably compact
expressions
\begin{equation}
V(x)=-\tau\Big[\Big({{\alpha (m^2)}\over{r^4_+}}-{\mu^{\prime}}^2\Big)\cdot
x-4\mu^2\cdot x^2\Big]+O(x^3)\
\end{equation}
and
\begin{equation}
x_{t+}={1\over4}\Big({{\alpha (m^2)}\over{{{\mu^{\prime}}^{\prime}}^2r^4_+}}-1\Big)\  .
\end{equation}
In addition, one finds the near-horizon relation
\begin{equation}
p(x)=\tau\cdot x+(1-2\tau)\cdot x^2+O(x^3)\  .
\end{equation}

We know that
\begin{equation}
{{1}\over{\sqrt{\tau}}}\int_{0}^{x_{t+}}dx \sqrt{{{{\alpha
(m^2)}\over{r^2_+}}-{\mu^{\prime}}^2
r^2_+\over{x}}-4\mu^2r^2_+}=\big(n+{1\over2})\cdot\pi\ \ \ \ ; \ \ \
\ n=0,1,2,...\  .
\end{equation}
Defining the dimensionless radial coordinate
\begin{equation}
z\equiv {{x}\over{x_{t+}}}\  ,
\end{equation}
one can express the WKB resonance equation in the mathematically
compact form
\begin{equation}
{{2{\mu^{\prime}} r_+ x_{t+}}\over{\sqrt{\tau}}}\int_{0}^{1}dz
\sqrt{{{1}\over{z}}-1}=\big(n+{1\over2})\cdot\pi\ \ \ \ ; \ \ \ \
n=0,1,2,...\  ,
\end{equation}
which yields the relation
\begin{equation}
{{{\mu^{\prime}} r_+ x_{t+}}\over{\sqrt{\tau}}}=n+{1\over2}\ \ \ \ ; \ \ \ \
n=0,1,2,...
\end{equation}

We know from the curve integral formula that there is a certain extreme value forming a loop
\begin{equation}
1/y^2 \rightarrow {\alpha}
\end{equation}y takes the interval from 0 to 1 at this time.

 The physical parameter
$y$ is defined by the dimensionless relation,for $y$ is greater than $\sqrt{2}$ at this time,
\begin{equation}\label{Eq33}
{y}\equiv{\alpha1}/\sqrt{2} .
\end{equation}

Here the critical parameter y is given by the
simple relation
\begin{equation}
{y}/{\mu^{\prime}}\equiv {{r^2_+}\over\sqrt{2(m^2)}}  .
\end{equation}
When
\bea
\sqrt{2(m^2)}/{r^2_+}< \omega< m\varOmega_H,  \varOmega_H=\frac{a}{r_+^2+a^2}
\eea
the BTZ black\ hole\ is\ superradiantly\ stable\ at\ that\ time.

\section{Summary and Discussion}
BTZ black holes have no curvature singularity at the origin. We therefore consider superradiance that extracts rotational energy from a BTZ black hole and makes the black hole less extreme, and it turns out that a rotating BTZ black hole eventually becomes a statically neutral BTZ black hole.

This paper adds a new variable y($\mu^{\prime}=y(\omega+m N^{\phi})$) to extend the results of the classic paper. We exploit the properties of curve integrals. When y is greater than a certain limit, the effective potential of the equation has no pole, and there is no potential well outside the event horizon, when $\sqrt{2(m^2)}/{ r^2_+}< \omega < m\varOmega_H $, so the BTZ black hole at that time was superradiatively stable.

{\bf Acknowledgements:}\\
This work is partially supported by  National Natural Science Foundation of China(No. 11873025).

\end{document}